\documentclass[12pt]{article}

\usepackage{amsmath,amssymb}
\usepackage{feynmp}
\usepackage{graphicx}

\newcommand{\mot}{\!\not \!}

\def\tr {{\rm Tr}}
\def\e {{\rm e}}
\def\c {\cos}
\def\s {\sin}

\def\mot {\not \!}
\def\iy {\int_{-\pi R}^{\pi R}\!\!\!\!{\rm d}y}
\def\dk{\int\!\!\frac{d^4k}{(2\pi)^4i}}         
\def\dx{\int_0^1\!\!dxdy~}

\def\nt{\notag}                                

\def\dk4{\int\!\!\frac{{\rm d}^4k}{(2\pi)^4i}}	%
\def\dx{\int_0^1\!\!\!\!{\rm d}x\!\!\int_0^{1-x}\!\!\!\!\!\!\!\!{\rm d}y~}%
\def\e{{\rm e}}	%
\def\nt{\notag}					%
\makeatletter
    
    \@addtoreset{equation}{section}
\makeatother

\begin{document}
\setlength{\baselineskip}{18pt}
\begin{titlepage}

\begin{flushright}
KOBE-TH-09-02
\end{flushright}
\vspace{1.0cm}
\begin{center}
{\LARGE\bf Neutron Electric Dipole Moment \\
\vspace*{5mm}
in the Gauge-Higgs Unification} 
\end{center}
\vspace{25mm}

\begin{center}
{\large
Yuki Adachi, 
%\footnote{e-mail : yuki1983@kobe-u.ac.jp},
%
C. S. Lim
%\footnote{e-mail : lim@kobe-u.ac.jp}
%
%
and Nobuhito Maru$^*$
}
\end{center}
\vspace{1cm}
\centerline{{\it Department of Physics, Kobe University,
Kobe 657-8501, Japan}}

\centerline{{\it
$^*$Department of Physics, Chuo University, 
Tokyo 112-8551, Japan
}}
%
%   Abstract
%
\vspace{2cm}
\centerline{\large\bf Abstract}
\vspace{0.5cm}
We study the neutron electric dipole moment (EDM) in a five dimensional 
$SU(3)$ gauge-Higgs unification compactified on $M^4\times S^1/Z_2$ space-time
including a massive fermion. 
We point out that to realize the CP violation is a non-trivial task in the gauge-Higgs unification scenario 
and argue how the CP symmetry is broken spontaneously by the VEV of the Higgs, 
the extra space component of the gauge field. 
We emphasize the importance of the interplay between the VEV of the Higgs and the $Z_{2}$-odd bulk mass term 
to get physically the CP violation. 
We then calculate the one-loop contributions to the neutron EDM as the typical example of the CP violating observable 
and find that the EDM appears already at the one-loop level, without invoking to the three generation scheme. 
We then derive a lower bound for the compactification scale, 
which is around 2.6 TeV, by comparing the contribution due to the nonzero Kaluza-Klein modes with the experimental data. 
\end{titlepage} 

\newpage

\section{Introduction}

Gauge-Higgs unification scenario proposed long time ago \cite{Manton, Fairlie, Hosotani,Antoniadis} 
has attracted recent revived interest as one of the attractive scenarios 
solving the hierarchy problem without invoking supersymmetry. 
In this scenario, Higgs doublet in the Standard Model (SM) is identified with 
the extra spatial components of the higher dimensional gauge fields. 
Remarkable feature is that the quantum correction to Higgs mass is finite and insensitive to the cutoff scale of the theory, 
in spite of the fact that higher dimensional gauge theories are generally non-renormalizable. 
The reason is simply that the Higgs mass-squared term as a local operator is forbidden by the higher dimensional gauge invariance. 
The radiatively induced finite Higgs mass should be understood as to be described by the Wilson line phase, 
that is a non-local operator and free from UV-divergence. 
This fact has opened up a new avenue to the solution of the hierarchy problem \cite{HIL}. 
Since then, much attention has been paid to the gauge-Higgs unification 
and many interesting works have been done from various points of view \cite{KLY}-\cite{HSY}. 

The finiteness of Higgs mass has been studied and verified in various models 
and types of compactification at one-loop level 
\cite{ABQ}-\cite{LMH}\footnote{For the case of gravity-gauge-Higgs unification, 
see \cite{HLM}} and even at two loop level \cite{MY}.
%, HMTY}. 
It is natural to ask whether any other finite calculable physical observables exist in the gauge-Higgs unification. 
In a paper by the present authors \cite{ALM1}, we have found a striking fact:  
we have shown that the anomalous magnetic moment of fermion in the $(D+1)$ dimensional QED gauge-Higgs unification model 
compactified on $S^1$ becomes finite for an arbitrary space-time dimension. 
The reason is easily understood relying on an operator analysis. 
In four dimensional space-time, 
a dimension six gauge invariant local operator describes the anomalous magnetic moment: 
\begin{align}
i \bar{\psi}_L \sigma^{\mu\nu} \psi_R F_{\mu\nu} \langle H \rangle. 
\label{MMM4}
\end{align}
However, when included into the scheme of gauge-Higgs unification, the Higgs doublet should be replaced 
by an extra space component of the higher dimensional gauge field $A_y$. 
Then, to preserve the gauge symmetry, $A_y$ should be further replaced by gauge covariant derivative $D_y$, 
and the relevant gauge invariant operator becomes 
\begin{align}
i \bar{\Psi} \Gamma^{MN} D_L \Gamma^L \Psi F_{MN}  
\label{MMMD}
\end{align}
where $L, M$ and $N$ denote $(D+1)$ dimensional Lorentz indices. 
The key observation of our argument is that the operator (\ref{MMMD}), 
when $D_L$ is replaced by $\langle D_L \rangle$ with the gauge field $A_L$ replaced by its VEV, vanishes 
because of the on-shell condition $i \langle D_L \rangle \Gamma^L \Psi = 0$. 
As the local operator is forbidden, the anomalous magnetic moment is expected to be free from the UV-divergence.  
We confirmed the finiteness of the magnetic moment by an explicit diagrammatical calculations \cite{ALM1}. 
This is a remarkable specific prediction of the gauge-Higgs unification to be contrasted with 
the case of Randall-Sundrum model \cite{g-2RS} or the universal extra dimension scenario \cite{g-2UED}, 
in which the magnetic moment of fermion diverges in the models with more than five space-time dimensions. 

Although this result was quite impressive, the above model is too simple to be realistic. 
In particular, the famous result by Schwinger in ordinary QED could not be reproduced 
as the contribution of zero-modes in the simplified model. 
Thus, in our following paper \cite{ALM2}, 
we have clarified the issue on cancellation mechanism of ultraviolet (UV) divergences in a realistic gauge-Higgs unification model. 
What we adopted was  $(D+1)$ dimensional $SU(3)$ gauge-Higgs unification model 
compactified on an orbifold $S^1/Z_2$ with a massive bulk fermion in a fundamental representation, 
whose gauge group is large enough to incorporate that of the Standard Model. 
The orbifolding is indispensable to obtain chiral theory and to reduce the gauge symmetry to that of the Standard Model. 
In order to obtain a realistic Yukawa coupling we introduced a bulk mass parameter of fermion, 
which should have odd $Z_2$ parity in order to preserve the $Z_2$ symmetry. 
The bulk mass causes localization of fermions with different chiralities at different fixed points of the orbifold. 
Hence the overlap integral of their mode functions yields an exponentially suppressed Yukawa coupling. 
In this way, we can freely obtain the light fermion masses, which are otherwise of ${\cal O}(M_W)$ 
in the gauge-Higgs unification scenario, by tuning the bulk mass parameters. 
We thus have succeeded in recovering the Schwinger's result, 
still keeping the nice feature of the scenario, i.e. the anomalous moment was shown to be finite even in 6 dimensional space-time, 
where other higher dimensional theories such as universal extra dimension scenario give divergent results. 
In the most recent paper \cite{ALM3}, 
we also have performed numerical calculations to obtain the contribution of non-zero KK modes 
to the muon anomalous magnetic moment and have derived a useful constraint on the compactification scale 
by comparing the result with the experimental data.   

In this paper, we focus on the CP violation in the gauge-Higgs unification scenario. 
As the concrete example of the physical observable due to the CP violation 
we discuss the neutron electric dipole moment (EDM) whose computation has some similarity 
to that of the anomalous magnetic moment of fermions. 
We will work in the  same model as in the previous paper \cite{ALM3}, 
i.e. the 5 dimensional $SU(3)$ gauge-Higgs unification model 
compactified on an orbifold $S^1/Z_2$ with a massive bulk fermion in a fundamental representation. 

Let us note that how to break CP symmetry is a non-trivial question in the gauge-Higgs unification scenario, 
since the Higgs field is nothing but a gauge field to start with and its Yukawa coupling is originally gauge coupling, which is real. 
As far as the theory itself has CP symmetry, 
the possible way to break CP is due to the compactification which does not respect the symmetry 
as in the case of Calabi-Yau manifold with non-trivial complex structure \cite{Lim} 
or by the VEV of some field which has odd CP eigenvalue \cite{Poland}. 
Both mechanisms may be understood as (a sort of) spontaneous CP violation, 
since the theory itself preserves the CP symmetry and 
the way of the compactification is responsible for the determination of the vacuum state. 
(In fact, the effect of compactification is accompanied by the compactification scale $1/R$, 
which has a mass dimension and the corresponding CP violation is ``soft".)     
 
In the present model the compactification itself is too simple to break CP, 
since the orbifold is trivially invariant under a discrete transformation $y \to -y \ (y: \mbox{extra space coordinate})$. 
Thus the possible unique source to break the CP symmetry is expected to be the VEV of the Higgs field, 
which is the zero-mode of $A_{y}$, the extra space component of the gauge field.    

To see whether this is really the case or not, 
we argue how the space-time coordinates and each field behave under the P and CP transformations. 
First, let us note that the EDM is P- and CP-odd observable, 
and therefore both of P and CP have to be broken to get a non-vanishing EDM. 
The P and CP transformations in higher dimensional theories need some care. 
Though we can easily find P and C transformations in higher dimensional sense, 
they may not reduce to ordinary 4-dimensional P or C transformations 
when dimensional reduction is performed \cite{Lim}. 
In the 5 dimensional space-time, however, 
the spinor is a 4-component one just as in the 4-dimensional theory 
and P and C transformations may be defined in the ordinary ways. 

First the parity transformation is defined for fermions as  
\begin{equation} 
\label{Ptransf} 
P: \ \ \Psi \to \gamma^{0} \Psi, 
\end{equation}
where $\Psi$ denotes the $SU(3)$ triplet fermion.
To be precise, the extra-space coordinate $y$ turns out to be enforced to change
its sign for the kinetic term to be invariant under (\ref{Ptransf})
and at the first glance it does not seem to  correspond to the ordinary
4-dimensional P transformation. 
However, at least the zero-mode fields corresponding to the ordinary particles
in the Standard Model are even functions of $y$ 
and the change of the sign is irrelevant for the low-energy effective theory.
Let us note that in our model the P symmetry is broken anyway by the orbifolding
, no matter $A_{y}$ develops its VEV or not, 
since the orbifolding is aimed to realize a chiral theory.
This may also be known by realizing that the orbifold condition for the fermion 
\begin{equation}
\Psi(- y)
= P \gamma^5 \Psi(y) \ \ \ (P={\rm diag}(+,+,-)) 
\label{orbifolding}
\end{equation}
is inconsistent with the parity transformation (\ref{Ptransf}), since $\gamma^{0}$ does not commute with $\gamma^{5}$.

Next, combining with the C transformation, $C: \ \ \Psi \to i \gamma^{2} \Psi^{\ast}$, we can derive the CP transformation:  
\begin{equation} 
\label{CPtransf} 
CP: \ \ \Psi (x^{\mu}, y) \to i\gamma^{0} \gamma^{2} \Psi (x_{\mu}, y)^{\ast},     
\end{equation}
This time, the transformation is consistent with the condition (\ref{orbifolding}), since $\gamma^{0} \gamma^{2}$ commutes with $\gamma^{5}$. Hence, CP is not violated by the orbifolding. The corresponding transformation properties of the space-time coordinates 
and the gauge field are fixed so that $\bar{\Psi}i \Gamma^M (\partial_M -igA_M)\Psi \ 
(\Gamma^{M} = (\gamma^{\mu}, i\gamma^{5}), \ A_M = (A_{\mu}, A_{y}) \ (\mu = 0-3))$ is invariant under (\ref{CPtransf}). Namely,  
\begin{equation} 
\label{CPcoordinatetransf} 
CP: \ \ x^{\mu} \to x_{\mu}, \ y \to y, \ A_{\mu}(x^{\mu}, y) \to - A^{\mu}(x_{\mu}, y)^{t}, \ A_{y}(x^{\mu}, y) \to - A_{y}(x_{\mu}, y)^{t}.    
\end{equation}
The $Z_{2}$-odd bulk mass term in the lagrangian $- M \epsilon(y) \bar{\Psi} \Psi \ \ (\epsilon(y): \mbox{sign function of}~y)$ is also invariant under such defined CP transformation, as $y$ remains untouched and $\epsilon (y)$ does not change its sign.  Let us note that if the fermions are expanded in terms of the ortho-normal set of plane waves as $\Psi (x^{\mu}, y) = \sum_{n} e^{i \frac{n}{R}y} \Psi^{(n)} (x^{\mu})$ with $R$ being the radius of the circle (though real mass eigenstates have different mode functions in the presence of the bulk mass $M$), the CP transformation necessitates the exchange of the KK modes, $n \leftrightarrow -n$, in addition to the 4-dimensional CP transformation 
for $\Psi^{(n)} (x^{\mu})$. Fortunately, this exchange of the KK modes is irrelevant for the zero-mode fermions. Thus the transformation given in (\ref{CPtransf}) and (\ref{CPcoordinatetransf}) just reduce to the ordinary 4-dimensional CP transformation for zero-mode fields.  

We thus realize that $A_{y}$ has CP eigenvalue $-1$. Hence, the VEV of $A_{y}$ is the unique source of the CP violation. As the matter of fact, however, in the case that the $Z_{2}$-odd bulk mass term vanishes, the CP violation is known to disappear even for the non-vanishing VEV of $A_{y}$. 
In fact, in this case, we can perform a chiral transformation, $\Psi \to e^{i\frac{\pi}{4}\gamma^{5}} \Psi$ such that $i\gamma^{5}$ disappears from the covariant derivative term $i\bar{\Psi} \Gamma^{5} D_{5} \Psi$, keeping the other parts of the lagrangian invariant.  
Now, $A_{y}$ has a scalar type coupling with fermions and 
therefore is now even under the CP transformation: 
\begin{eqnarray} 
\label{CPcoordinatetransf2} 
CP: && \ \ x^{\mu} \to x_{\mu}, \ y \to y, \ \Psi (x^{\mu}, y) \to i\gamma^{0} \gamma^{2} \Psi (x_{\mu}, -y)^{\ast}, \nonumber \\ 
&&  \ A_{\mu}(x^{\mu}, y) \to - A^{\mu}(x_{\mu}, -y)^{t}, \ A_{y}(x^{\mu}, y) \to A_{y}(x_{\mu}, -y)^{t}. 
\end{eqnarray}
The invariance of the action under the CP transformation is easily checked by use of the change of the integration variable $y \to -y$. Thus the VEV of $A_{y}$ no longer violates CP. Let us note that in this case the exchange of the KK modes is not needed for fermions.  

We thus find that to break CP physically and to get a non-vanishing EDM, 
the interplay between the VEV of $A_{y}$ and the bulk mass $M$ is crucial, 
and from such a point of view both of the VEV and the bulk mass are the cause of the CP violation on an equal footing. 
The necessity of the interplay will be shown by an explicit calculation of Feynman diagrams later in this paper. 

Let us note that the VEV of $A_{y}$ is needed anyway to get the EDM, 
since the gauge invariant operator to describe the EDM in the standard model is 
\begin{equation} 
-\frac{i}{2}\bar{\psi} \sigma^{\mu\nu} \gamma^5 F_{\mu\nu} \langle H \rangle \psi,  
\end{equation}
which vanishes when $\langle H \rangle$ and therefore the VEV of $A_{y}$ vanishes. 
From the same reasoning to conclude that the anomalous magnetic moment is finite 
even for 6-dimensional space-time, 
we expect relying on a similar operator analysis that the EDM is also finite even for 6D theory, 
though in this paper we work in the 5D space-time.

The purposes of this paper are two-folds. 
One is to confirm that the EDM really appears as a finite calculable observable already at the one-loop level, 
though the EDM has been shown to appear only at the three loop level in the Standard Model \cite{Shabalin}. 
In addition, in our model we introduce only the first generation and to get the EDM 
we do not need 3 generation scheme, in clear contrast to the case of the Standard Model. 
The other one is to obtain the lower bound on the compactification scale, i.e. 
the upper bound on the size of the extra space, by comparing the prediction of our model with the experimental data.     
        
In ref.\cite{Poland}, CP violation in the gauge-Higgs unification scenario has also been discussed. 
The gauge group they adopt is a $U(1)$ and the extra space is a circle. 
In our case, the gauge group is an $SU(3)$ and the extra space is orbifold 
so that the model can incorporate the chiral theory of the Standard Model. 
The chiral theory clearly violates the P symmetry, 
in contrast to the case of $U(1)$ gauge theory discussed in \cite{Poland}. 
The introduction of the $Z_{2}$-odd bulk mass term is also a new feature of our model. 
By adopting such realistic model, 
we hope that we can derive a realistic prediction for the neutron EDM to be compared with the data, 
which is expected to have various new ingredients not seen in the prediction in \cite{Poland}, 
due to the complexity of our model. 
Notice that the possibility of the recovery of CP symmetry 
due to the Wilson line phase $\pi$ at the minimum of the effective potential for $A_{y}$ pointed out in \cite{Poland} 
has no relevance in our model, as we assume the realistic situation where the weak scale, 
i.e. the VEV of $A_{y}$ times gauge coupling is much smaller than the compactification scale $1/R$. 

This paper is organized as follows. 
In the next section, 
we briefly summarize our model and discuss the mass eigenvalues and corresponding mode functions of fermions and gauge bosons. 
In section 3, we derive general formulae relevant for EDM 
concerning a few types of Feynman diagrams where 4D gauge boson $A_{\mu}$ or 4D scalar $A_{y}$ are exchanged 
or self-interaction of the 4D gauge and scalar fields is contributing. 
The coupling constants in the interaction vertices are left arbitrary there.  
Then combining with the interaction vertices derived in ref. \cite{ALM3} 
we obtain the contribution of each type of Feynman diagram to EDM. 
In section 4, we numerically estimate the contribution of nonzero KK modes to the EDM 
as the function of the compactification scale $1/R$. 
Comparing with the experimental data we finally obtain a rather stringent lower bound for the compactification scale. 
Section 5 is devoted to the summary discussion.

\section{The Model}

Since we employ the same model as that discussed in \cite{ALM2} to calculate EDM, 
we briefly summarize it in this paper. 
We consider a five dimensional $SU(3)$ gauge-Higgs unification model 
compactified on an orbifold $S^1/Z_2$ with a radius $R$ of $S^1$. 
As a matter field, 
a massive bulk fermion in the fundamental representation 
of $SU(3)$ gauge group is introduced.
The Lagrangian is given by
\begin{equation}
 {\cal L}
=
 -\frac{1}{2}\tr (F_{MN}F^{MN})+\bar{\Psi}(i D\!\!\!\!/ -M\epsilon(y))\Psi
\end{equation}
where the indices $M,N=0,1,2,3,5$, the five dimensional 
gamma matrices are $\Gamma^M=(\gamma^{\mu},i\gamma^{5})$
($\mu=0,1,2,3$),
\begin{align}
 F_{MN}
=&
 \partial_MA_N-\partial_NA_M -ig[A_M,A_N],
 \\
 \not \!\! D 
=&
 \Gamma^M (\partial_M -igA_M),
 \\
 \Psi
=&
 (\Psi_1,\Psi_2,\Psi_3)^{\rm T}
\end{align}
$g$ denotes a gauge coupling constant in five dimensional gauge theory.
$M$ is a bulk mass of the fermion. 
$\epsilon(y)$ is a sign function of an extra coordinate $y$ which is necessary 
to introduce a $Z_2$ odd bulk mass term. 

The periodic boundary condition is imposed along $S^1$ and $Z_2$ parity
assignments are taken as
\begin{equation}
%\begin{array}{c}
 A_{\mu}(y_i-y) = P A_\mu(y_i+y) P^\dag, \quad 
% \left(
% \begin{array}{ccc}
%  (+,+)&(+,+)&(-,-)\\
%  (+,+)&(+,+)&(-,-)\\
%  (-,-)&(-,-)&(+,+)
% \end{array}
% \right),
% ~~
A_y(y_i-y) = -P A_y(y_i+y) P^\dag, \quad 
% \left(
% \begin{array}{ccc}
%  (-,-)&(-,-)&(+,+)\\
%  (-,-)&(-,-)&(+,+)\\
%  (+,+)&(+,+)&(-,-)
% \end{array}
% \right),
% \\
 \Psi(y_i-y)
= P \gamma^5 \Psi(y_i+y)
% \left(
% \begin{array}{c}
%  \Psi_{1L}(+,+) + \Psi_{1R}(-,-)\\
%  \Psi_{2L}(+,+) + \Psi_{2R}(-,-)\\
%  \Psi_{3L}(-,-) + \Psi_{3R}(+,+) 
% \end{array}
% \right)
%\end{array}
\label{BC}
\end{equation}
where $P={\rm diag}(+,+,-)$ at fixed points $y_i=0, \pi R$.
By this $Z_2$ parity assignment, $SU(3)$ is explicitly broken to $SU(2) \times U(1)$. 
Higgs scalar field is identified with the off-diagonal block of zero mode $A_y^{(0)}$.  

%%%%%%%%%%%%%%%%%%%%%%%%%%%%%%%%%%%%%%%%%%%%%%%%%%%%%%%%%%%%%%%%%%%%%%%%
%\subsection{The mass eigenvalues and mode functions of gauge boson}

The 4-dimensional gauge bosons $A_{\mu}$ and their scalar partners $A_y$
can be expanded in KK modes such that the boundary conditions (\ref{BC}) are satisfied, 
\begin{align}
 A_{\mu,y}(x,y)
=& \frac{1}{\sqrt{2\pi R}} A_\mu^{(0)}(x) 
+ \frac{1}{\sqrt{\pi R}} \sum_{n=1}^{\infty} A_{\mu,y}^{(n)}(x) \cos \left( \frac{n}{R}y \right)~({\rm even}), \\
 A_{\mu,y}(x,y)
=&
 \frac{1}{\sqrt{\pi R}} \sum_{n=1}^{\infty} A_{\mu,y}^{(n)}(x) \sin \left( \frac{n}{R}y \right)~({\rm odd}). 
\end{align}
After electroweak symmetry breaking, 
quadratic terms relevant to the gauge boson mass are diagonalized as
\begin{equation}
\begin{aligned}
 &\mathcal{L}_{\rm mass} - \frac{1}{2}\iy
 [\partial^{\mu}A_{\mu}^a - (\partial_y A_{y}^a-2m_W f^{6ab}A_{by})]^2
 \\
=&
 \sum_{n=1}^{\infty}
 \Bigg[
  \frac{1}{2}M_n^2 (\gamma_{\mu}^{(n)}\gamma^{\mu(n)}+h_{\mu}^{(n)}h^{\mu (n)})
  +\frac{1}{2}(M_n-2m_W)^2\phi_{\mu}^{(n)}\phi^{\mu(n)}
  +\frac{1}{2}(M_n+2m_W)^2Z_{\mu}^{(n)}Z^{\mu(n)}
  \\
&
  +(M_n+m_W)^2W_{\mu}^{+(n)}W^{-\mu(n)}
  +(M_n-m_W)^2X_{\mu}^{+(n)}X^{-\mu(n)}
 \Bigg]
  +\frac{1}{2}(2m_W)^2Z_{\mu}Z^{\mu}+m_W^2W^+_{\mu}W^{-\mu}
\\
&
 - \sum_{n=1}^{\infty}
 \Bigg[
  \frac{1}{2}M_n^2 (\gamma_y^{(n)}\gamma^{(n)}_y+h_y^{(n)}h^{(n)}_y)
  +\frac{1}{2}(M_n+2m_W)^2\phi_y^{(n)}\phi^{(n)}_y
  +\frac{1}{2}(M_n-2m_W)^2Z_y^{(n)}Z^{(n)}_y
  \\
&
  +(M_n-m_W)^2W_y^{+(n)}W^{-(n)}_y
  +(M_n+m_W)^2X_y^{+(n)}X^{-(n)}_y
 \Bigg]
  +\frac{1}{2}(2m_W)^2Z_yZ_y+m_W^2W^+_yW^-_y
\end{aligned}
\end{equation}
where the gauge-fixing term in 't Hooft-Feynman gauge is introduced 
to eliminate the mixing terms between the gauge bosons and the gauge scalar bosons. 
%The 't Hooft-Feynman gauge $\xi=1$ is adopted throughout this paper. 
$M_n=\frac{n}{R},m_W=2g\langle A_y^6\rangle = \frac{g}{\sqrt{2\pi R}}v=g_4 v$. 
$g_4$ is a four dimensional gauge coupling. 
The KK mass eigenstates and zero mode mass eigenstates are given by 
\begin{equation}
\begin{array}{ll}
\gamma^{(n)}
=\frac{1}{2}(\sqrt{3}A^{3(n)} + A^{8(n)}),
&h^{(n)}=A^{6(n)}, 
\\ 
Z^{(n)}=\frac{1}{\sqrt{2}}
 \left[\frac{A^{3(n)} - \sqrt{3}A^{8(n)}}{2} -A^{7(n)} \right],
&\phi^{(n)}=\frac{1}{\sqrt{2}}
 \left[\frac{A^{3(n)} - \sqrt{3}A^{8(n)}}{2} + A^{7(n)} \right],
\\
W^{\pm(n)}=\frac{1}{2}
 \left[A^{1(n)} + A^{5(n)} \mp i(A^{2(n)} - A^{4(n)}) \right],
&X^{\pm(n)}
=\frac{1}{2}
 \left[A^{2(n)} + A^{4(n)} \mp i (-A^{1(n)} + A^{5(n)}) \right],
\\
\gamma_{\mu}=\frac{1}{2}\left[\sqrt{3}A^3_{\mu} + A^8_{\mu} \right], 
&h=A_{y}^{6(0)}, 
\\
W^{\pm}_{\mu}=\frac{1}{\sqrt{2}}(A^1_{\mu} \mp i A^2_{\mu}),
&X^{\pm}=\frac{1}{\sqrt{2}}\left[A^4_{y}\mp i A^5_{y} \right], 
\\
Z_{\mu}=\frac{1}{2}(A^3_{\mu}-\sqrt{3}A^8_{\mu}), 
&\phi=A^7_{y}. 
\end{array}
\end{equation}
The zero mode gauge bosons $W^{\pm}_{\mu},Z_{\mu},\gamma_{\mu}$
correspond to $W$ boson, $Z$ boson and photon and zero mode scalar fields 
$X^{\pm},\phi,h$ correspond to charged NG boson, neutral NG boson,
Higgs field in the Standard Model, respectively.

Some comments on this model are in order. 
First, the predicted Weinberg angle of this model is not realistic, $\sin^2 \theta_W = 3/4$
\cite{Wudka}
. 
Possible way to cure the problem is to introduce an extra $U(1)$ or 
the brane localized gauge kinetic term \cite{SSS}. 
Second, the up quark remains massless and we have no up-type Yukawa coupling. 
A possible way out of this situation is to introduce second-rank symmetric tensors of $SU(3)$ 
(${\bf 6}$ dimensional representation) \cite{CCP}. 

On the other hand, 
we have obtained a quadratic part of 4D effective Lagrangian of fermion
\begin{equation}
\begin{aligned}
 \mathcal{L}
=&
 \sum_{n=1}^{\infty}
 \left[
  \bar{\psi}_1^{(n)}(i \partial\!\!\!/ -m_n)\psi_1^{(n)}
  +\bar{\psi}_2^{(n)}(i \partial\!\!\!/ -m_n^-)\psi_2^{(n)}
  +\bar{\psi}_3^{(n)}(i\partial\!\!\!/ -m_n^+)\psi_3^{(n)}
 \right]
%\\&
+\bar{d}(i \partial\!\!\!/ -m)d + \bar{u}_L i \partial\!\!\!/ u_L. 
\end{aligned}
\end{equation}
where the mass eigenstates of fermion were obtained as:
\begin{align}
 d_L
=& 
 \Psi^{(0)}_{2L}+\sum_{n=1}^{\infty} \frac{\hat{m}_n}{m_n}\Psi^{(n)}_{3L}
  ,~~~
  d_R=\Psi^{(0)}_{3R}
 +\sum_{n=1}^{\infty}(-1)^n\frac{\hat{m}_n}{m_n}\Psi^{(n)}_{3R},
\\
 \psi_{3L}^{(n)}
=&
 \frac{1}{\sqrt{2}}\left[
 \Psi_{2L}^{(n)}+\Psi_{3L}^{(n)}
 +\frac{M^2}{2m_n^3}m_W(\Psi_{2L}^{(n)}-\Psi_{3L}^{(n)})
 -\frac{\hat{m}_n}{m_n}\Psi_{2L}^{(0)}
 +\sum_{l\neq n}^{\infty} \frac{\tilde{m}_{nl}}{m_n^2-m_l^2}
  (m_l\Psi^{(l)}_{3L}-m_n\Psi_{2L}^{(l)})
 \right],
 \\
 \psi_{2L}^{(n)}
=&
 \frac{1}{\sqrt{2}}\left[
 \Psi_{2L}^{(n)}-\Psi_{3L}^{(n)}
 -\frac{M^2}{2m_n^3}m_W(\Psi_{2L}^{(n)}-\Psi_{3L}^{(n)})
 +\frac{\hat{m}_n}{m_n}\Psi_{2L}^{(0)}
 +\sum_{l\neq n}^{\infty} \frac{\tilde{m}_{nl}}{m_n^2-m_l^2}
  (m_l\Psi^{(l)}_{3L}+m_n\Psi_{2L}^{(l)})
 \right],
 \\
 \psi_{3R}^{(n)}
=&
 \frac{1}{\sqrt{2}}\left[
 \Psi_{2R}^{(n)}+\Psi_{3R}^{(n)}
 -\frac{M^2}{2m_n^3}m_W(\Psi_{2L}^{(n)}-\Psi_{3L}^{(n)})
 -(-1)^n\frac{\hat{m}_n}{m_n}\Psi_{2L}^{(0)}
 +\sum_{l\neq n}^{\infty} \frac{\tilde{m}_{nl}}{m_n^2-m_l^2}
  (m_n\Psi^{(l)}_{3R}-m_l\Psi_{2R}^{(l)})
 \right],
 \\
 \psi_{2R}^{(n)}
=&
 \frac{1}{\sqrt{2}}\left[
 \Psi_{2R}^{(n)}-\Psi_{3R}^{(n)}
 -\frac{M^2}{2m_n^3}m_W(\Psi_{2R}^{(n)}-\Psi_{3R}^{(n)})
 +\frac{\hat{m}_n}{m_n}\Psi_{2R}^{(0)}
 +\sum_{l\neq n}^{\infty} \frac{\tilde{m}_{nl}}{m_n^2-m_l^2}
  (m_l\Psi^{(l)}_{3R}+m_n\Psi_{2R}^{(l)})
 \right],
 \\
u_L=& \Psi_{1L}^{(0)}, \quad \psi_1^{(n)}=\Psi_1^{(n)}. 
% \\
%M_{{\rm diag}}=&(\mu,m_1^-,m_1^+,\cdots)
\end{align}
where
\begin{equation}
\begin{array}{ll}
m = \frac{2\pi RM}{\sqrt{(1-\e^{-2\pi RM})(\e^{2\pi RM}-1)}}m_W,
&\tilde{m}_{Wn}=\left(1-\frac{2M^2}{m_n^2}\right)m_W, 
\\
\hat{m}_n=4\sqrt{\frac{\pi RM}{1-\e^{-2\pi RM}}}
          \frac{1-(-1)^n\e^{-\pi RM}}{\pi Rm_n^3}M_n M m_W,
&\tilde{m}_n=(-1)^n\hat{m}_n,
\\
\tilde{m}_{nl}=\frac{4nl(1-(-1)^{n+l})}{\pi Rm_nm_l(n^2-l^2)}(1-\delta_{nl})m_W M, 
&(m_n^\pm)^2 = m_n^2 \pm 2m_W \frac{M_n^2}{m_n}. 
\end{array}
\label{mass}
\end{equation}
In deriving the above 4D effective Lagrangian of fermion, 
the following mode expansions are substituted and integrated out over the fifth coordinate. 
%%%%%%%%%%%%%%%%%%%%%%%%%%%%%%%%%%%%%%%%%%%%%%%%%%%%%%%%%%%%%%%%%%%%%%%%
%Expanding a bulk fermion $\Psi(x,y)$ in terms of parity even (odd) mode functions 
%$f_L^{(n)}, f_L^{(0)},f_R^{(n)} (g^{(n)})$, 
\begin{equation}
 \Psi(x,y)
=
 \sum_{n=1}^{\infty}
 \left(
 \begin{aligned}
  \Psi^{(n)}_{1L}(x)f_L^{(n)}(y)
   +\Psi^{(n)}_{1R}(x)g^{(n)}(y)\\
  \Psi^{(n)}_{2L}(x)f_L^{(n)}(y)
   +\Psi^{(n)}_{2R}(x)g^{(n)}(y)\\
  \Psi^{(n)}_{3L}(x)g^{(n)}(y)
   +\Psi^{(n)}_{3R}(x)f_R^{(n)}(y)
 \end{aligned}
 \right)
+
 \left(
 \begin{aligned}
  \Psi^{(0)}_{1L}(x)f_L^{(0)}(y)\\
  \Psi^{(0)}_{2L}(x)f_L^{(0)}(y)\\
  \Psi^{(0)}_{3R}(x)f_R^{(0)}(y)
 \end{aligned}
 \right)
\end{equation}
with the zero mode wave functions
\begin{equation}
 f^{(0)}_L=\sqrt{\frac{M}{1-\e ^{-2\pi RM}}}\e^{-M|y|}
 ~~,~~
 f^{(0)}_R=\sqrt{\frac{M}{\e^{2\pi RM}-1}}\e^{M|y|}.
\end{equation}
and the nonzero KK mode functions
\begin{align}
 f_L^{(n)}
=&
 \frac{M_n}{\sqrt{\pi R}m_n}
  \left[\c \left( \frac{n}{R}y \right) - \frac{MR}{n} \epsilon(y)\s \left( \frac{n}{R}y \right) \right],
 \\
 f_R^{(n)}
=&
 \frac{M_n}{\sqrt{\pi R}m_n}
  \left[ \c \left( \frac{n}{R}y \right) + \frac{MR}{n}\epsilon(y)\s \left( \frac{n}{R}y\right) \right],
 \\
 g^{(n)}
=&
 \frac{1}{\sqrt{\pi R}}\s \left( \frac{n}{R}y \right).
\end{align}
Deriving the vertex functions necessary for calculating the neutron EDM 
by using the above mass eigenfunctions is straightforward, but complicated. 
We do not repeat here their derivation 
since the necessary vertex functions are the exactly same ones as summarized in Appendix A of 
our previous paper \cite{ALM3}, 
except that the muon $\mu$ should be replaced by the down quark $d$.

\section{Calculation of the electric dipole moment}

In this section, 
we provide general formulae to calculate fermion EDM.
Various types of diagrams contributing to the EDM are written down below. 
Fermion  electric dipole moment is described by dimension 6 operator
$-\frac{i}{2}\bar{\psi}_L(p')\sigma^{\nu\rho}\gamma^5F_{\nu\rho}
\langle H \rangle\psi_R(p)$. 
%where $\sigma_{\mu\nu}\equiv \frac{i}{2}[\gamma_{\mu},\gamma_{\nu}]$,
%$F_{\mu\nu}=\partial_{\mu}A_{\nu}-\partial_{\nu}A_{\mu}$,
%respectively.
In general, quantum corrections to the photon vertex
$-\frac{e}{3} \gamma_{\rho}\bar{\psi}(p')\gamma^{\rho}\psi(p)$ can be written
as
\begin{equation}
-i\frac{e}{3} \gamma_{\mu}\bar{\psi}(p')[\gamma^{\mu}+\Gamma^{\mu}]\psi(p)
\end{equation}
%And $\Gamma^{\mu}$ forms into following respect to Lorentz invariance and 
%current conservation:
where 
\begin{equation*}
\Gamma^{\mu}=a_{\psi}\frac{p^{\mu}+p'^{\mu}}{2m_{\psi}}+
\frac{d_{\psi}}{e/3}(p^{\mu}+p'^{\mu})\gamma_5. 
\end{equation*}
$a_{\psi}$ and $d_{\psi}$ stand for the anomalous magnetic moment and 
the electric dipole moment of $\psi$, respectively.
Since our interest in this paper is in the electric dipole moment, 
the terms proportional to $\gamma_5 (p_\mu+p'_\mu)$ must be extracted. 

We now derive general formulae for each type of Feynman diagram,
leaving the couplings in the interaction vertices arbitrary.
First, the gauge boson exchange diagram is given by
($a, b, c, d$ are generic coupling constants)

\begin{align}
 \begin{array}{c}
 \includegraphics[scale=0.8]{./graph/fig6.1}
 \end{array}
=&
 \dk4 (aL +bR)\gamma^{\nu}\frac{-1}{\mot k+\mot p' -m_n}
 (Q_{\psi}\gamma_{\mu})\frac{-1}{\mot k+\mot p -m_n}
 \gamma_{\nu}(cL+dR)\frac{1}{k^2-M_G^2}
 \nt\\
\supset&
 -Q_{\psi} \dk4 \dx\frac{4(ac-bd)(1-X)m_n}
 {[k^2+X(1-X)m^2-Xm_n^2-(1-X)M_G^2]^3}
 \gamma_5P_{\mu}.  
\end{align}
In the second line only the part relevant for the EDM has been extracted.  
Similarly the diagram due to the exchange of the scalar partner of gauge boson
given by
\begin{align}
 \begin{array}{c}
 \includegraphics[scale=0.8]{./graph/fig7.1}
 \end{array}
=&
 \dk4 (aL+bR)\frac{-1}{\mot k+\mot p' -m_n}(Q_{\psi}\gamma_{\mu})
 \frac{-1}{\mot k +\mot p -m_n}(cL+dR)\frac{-1}{k^2-m_G^2}
 \nt\\
\supset&
 -Q_{\psi}\dk4\dx
 \frac{-(ac-bd)Xm_n}
 {[k^2-X^2m^2+Xm^2-Xm_n^2-(1-X)M_G^2]^3}
 \gamma_5P^{\mu}. 
\end{align}
For the diagrams due to the gauge boson self-energy, 
there are following three types of diagrams.
\begin{align}
\label{AmuAmu}
 \begin{array}{c}
 \includegraphics[scale=0.8]{./graph/fig8.1}
 \end{array}
=&
 \dk4 (aL+bR)\gamma_{\rho}\frac{-1}{\mot k-m_n}\gamma_{\nu}(cL+dR)
 \frac{1}{(k-p')^2-M_G^2}\frac{1}{(k-p)^2-M_G^2}
 \nt\\
 &\times(-e)[(2p'-p-k)_{\rho}\eta_{\mu\nu}+(2k-p'-p)_{\mu}\eta_{\nu\rho}+(2p-p'-k)_{\nu}\eta_{\mu\rho}]
 \nt\\
\supset&
 e\dk4 \dx \frac{-(ac-bd)[X+2+4(1-X)]m_n}
 {[k^2+X(1-X)m^2-XM_G^2-(1-X)m_n^2]^3}
 \gamma_5P^{\mu}, 
 \\
\label{AyAy}
 \begin{array}{c}
 \includegraphics[scale=0.8]{./graph/fig10.1}
 \end{array}
=&
 \dk4(aL+bR)\frac{-1}{\mot k -m_n}(cL+dR)\frac{-1}{(k-p')^2-M_G^2}
 \frac{-1}{(k-p)^2-M_G^2}e(2k-p'-p)_{\mu}
 \nt\\
\supset&
 -e\dk4\dx\frac{-(ac-bd)(1-X)m_n}
 {[k^2+X(1-X)m^2-XM_G^2-(1-X)m_n^2]^3}
 \gamma_5P_{\mu}, 
\end{align}
\begin{align}
\label{AmuAy}
 \begin{array}{c}
  \includegraphics[scale=.8]{./graph/fig12.1}
 \end{array}
 +({\rm h.c.})
=&
 \dk4(aL+bR)\gamma^{\mu}\frac{-1}{\mot k-m_n}(cL+dR)
 \frac{1}{(k-p')^2-M_G^2}\frac{-1}{(p-k)^2-M_G^2} \nt \\
& \times (\pm e)M_G 
 +({\rm h.c.})
 \nt\\
=&
 \pm e\dk4\dx \frac{(ac-bd)(x-y)M_G}{[k^2+X(1-X)m^2-X M_G^2-(1-X)m_n^2]^3}
 \gamma_5P^{\mu}
 \nt\\
=&
 0.  
\end{align}
Here, 
$M_G, m_n, m$ denote masses of the gauge boson, the internal fermion, and 
the down-type quark, respectively.
$Q_{\psi}$ denotes the electric charge of internal fermion.
$x, y$ are Feynman parameters and $X \equiv x+y$. 
$P_\mu$ is defined as the sum of the external momenta for fermions, $P_\mu \equiv p_\mu + p'_\mu$. 
In the last diagram of the gauge boson self-energy, 
the plus(minus) sign corresponds to the diagram 
where $W_\mu(X_\mu)$ boson propagates in the loop, respectively. 
In all amplitudes, we used the property that Feynman parameter integral of an odd function of $x-y$ vanishes. 

In order to arrive at the above expressions, 
the numerator of the integrand is calculated as 
\begin{align*}
& 
 (aL+bR)\gamma_{\nu}(\mot k+\mot p' +m_n)\gamma^{\mu}(\mot k+\mot p+m_n)
 \gamma^{\nu}(cL+dR)
 \\
\supset&
  - (ad-bc) \gamma_5 (x-y)(1-X)mP^{\mu}
 +2 (ac-bd) m_n\gamma_5 (1-X)P^{\mu}.
\end{align*}
In the above calculation, the momentum shift $k \to k-xp'-yp$ is performed 
and $\supset$ means that terms relevant for the fermion EDM are extracted. 
The equation of motion for the external fermion is also utilized
\begin{equation}
\bar{\psi}(p')\gamma_5(x \!\mot p'+y \!\mot p) \psi(p)
\to (-x+y)m\bar{\psi}(p')\gamma_5 \psi(p).
\end{equation}
Applying the possible interaction vertices described in \cite{ALM3} to these formulae derived above, 
we can obtain the amplitudes of the EDM in a straightforward way 
and list up them by classifying into the neutral current sector, charged current sector 
and gauge boson self-energy sector. 
Concerning the mode indices in the amplitude, the summation $\sum_{l,m,n=1}^\infty$ should be understood. 
In our calculation, we adopt approximations $m_W, m \ll 1/R$ and the results are shown 
at the leading order of ${\cal O}(m_W^2)$.

As will be seen below, all of the standard model diagrams have no contributions to the EDM at one-loop level,  
which is consistent with the well-known fact that the EDM in the standard model is generated 
at least at 3-loop level \cite{Shabalin}. 

\subsection{Neutral current sector}

\subsubsection{KK mode photon exchange}

\begin{equation}
 \begin{array}{c}
  \includegraphics[scale=0.7]{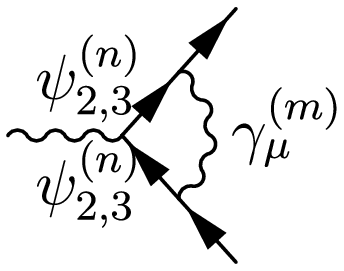}
%  \begin{picture}(0,0)(1,1)
%  \put(-42,12){$\psi_{2,3}^{(n)}$}
%  \put(-42,32){$\psi_{2,3}^{(n)}$}
%  \put(-5,22){$\gamma_{\mu}^{(m)}$}
%  \end{picture}
 \end{array}
\!\!=
 \frac{e}{3} \dk4 \dx 
 \frac{2\frac{g_4^2}{3}\frac{M_n^2}{m_n^2}
       \frac{\pi RM}{1-\e^{-2\pi RM}}\left(\frac{I_{3}}{\pi R}\right)^2
       \frac{m_l\tilde{m}_{nl}}{m_n^2-m_l^2}(-1)^{n+m}4(1-X)m_n}
 {[k^2-Xm_n^2-(1-X)M_m^2]^3}\gamma_5P_{\mu}. 
\end{equation}

\subsubsection{KK mode photon partner exchange}

\begin{equation}
 \begin{array}{c}
  \includegraphics[scale=0.7]{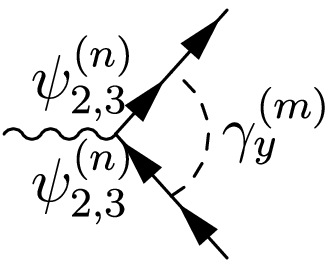}
%  \begin{picture}(0,0)(1,1)
%  \put(-42,12){$\psi_{2,3}^{(n)}$}
%  \put(-42,32){$\psi_{2,3}^{(n)}$}
%  \put(-5,22){$\gamma_y^{(m)}$}
%  \end{picture}
 \end{array}
=
 -\frac{e}{3} \dk4 \dx 
 \frac{2\frac{g_4^2}{3}\frac{\pi RM}{\e^{2\pi RM}-1}
       \left(\frac{I_{4}}{\pi R}\right)^2
       \frac{m_n\tilde{m}_{nl}}{m_n^2-m_l^2}(-1)^{n+m}Xm_n}
 {[k^2-Xm_n^2-(1-X)M_m^2]^3}\gamma_5P_{\mu}. 
\end{equation}

\subsubsection{Higgs exchange}

\begin{equation}
\begin{array}{c}
\includegraphics[scale=0.7]{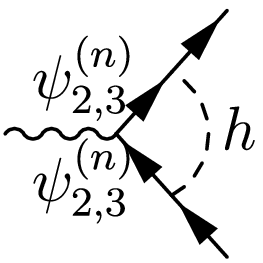}
%  \begin{picture}(0,0)(1,1)
%  \put(-42,12){$\psi_{2,3}^{(n)}$}
%  \put(-42,32){$\psi_{2,3}^{(n)}$}
%  \put(-5,22){$h$}
%  \end{picture}
\end{array}
=
 \frac{e}{3} \dk4\dx 
 \frac{g_4^2I_5^2\frac{m_l\tilde m_{nl}}{m_n^2-m_l^2}Xm_n(-1)^n}
 {[k^2-Xm_n^2]^3}\gamma_5P_{\mu}. 
\end{equation}

\subsubsection{KK mode Higgs exchange}

\begin{equation}
\begin{array}{c}
\includegraphics[scale=0.7]{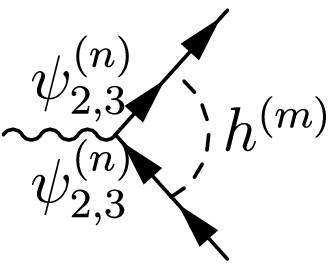}
%  \begin{picture}(0,0)(1,1)
%  \put(-42,12){$\psi_{2,3}^{(n)}$}
%  \put(-42,32){$\psi_{2,3}^{(n)}$}
%  \put(-5,22){$h^{(m)}$}
%  \end{picture}
\end{array}
=
 \frac{e}{3} \dk4\dx 
 \frac{2g_4^2 I_6^2(\frac{M_n}{\pi Rm_n})^2
       \frac{\pi RM}{\e ^{2\pi RM}-1}\frac{m_l\tilde{m}_{nl}}{m_n^2-m_l^2}
       (-1)^{n+m}Xm_n}
 {[k^2-Xm_n^2-(1-X)M_m^2]^3}\gamma_5P_{\mu}. 
\end{equation}

\subsubsection{KK mode Higgs partner exchange}

\begin{equation}
\begin{aligned}
\begin{array}{c}
\includegraphics[scale=0.7]{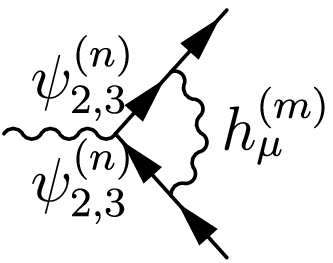}
%  \begin{picture}(0,0)(1,1)
%  \put(-42,12){$\psi_{2,3}^{(n)}$}
%  \put(-42,32){$\psi_{2,3}^{(n)}$}
%  \put(-5,22){$h^{(m)}_{\mu}$}
%  \end{picture}
\end{array}
=&
 -\frac{e}{3} \dk4\dx 
 \frac{8g_4^2 (1-X)m_n\gamma_5P_{\mu}}
 {[k^2-Xm_n^2-(1-X)M_m^2]^3}
 \\
&\times
 \Bigg[
 \frac{1}{2}\sqrt{\frac{\pi RM}{\rm e^{2\pi RM}-1}}\frac{\hat{m}_l}{m_l}
 \frac{I_{4}I_2}{\sqrt{\pi R}}
 ((-1)^{l+m+n}-1)
 -\frac{\pi RM}{\e^{2\pi RM}-1}\left(\frac{I_{4}}{\pi R}\right)^2
 \frac{m_n\tilde{m}_{nl}}{m_n^2-m_l^2}(-1)^{n+m}
 \Bigg]. 
 \end{aligned}
\end{equation}

\subsubsection{KK mode $\boldsymbol Z$ boson exchange}

\begin{equation}
\begin{aligned}
\begin{array}{c}
\includegraphics[scale=0.7]{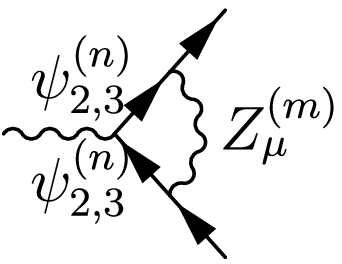}
%  \begin{picture}(0,0)(1,1)
%  \put(-42,12){$\psi_{2,3}^{(n)}$}
%  \put(-42,32){$\psi_{2,3}^{(n)}$}
%  \put(-5,22){$Z_{\mu}^{(m)}$}
%  \end{picture}
\end{array}
=&
 \frac{e}{3} \dk4\dx 
 \frac{g_4^2\frac{\pi RM}{\e^{2\pi RM}-1}\frac{m_n\tilde{m}_{nl}}{m_n^2-m_l^2}
 \left(\frac{I_{4}}{\pi R}\right)^2(-1)^{n+m}\left(\frac{M_m^2}{m_n^2}+1\right)4(1-X)m_n}
 {[k^2-Xm_n^2-(1-X)M_m^2]^3}\gamma_5P_{\mu}. 
 \end{aligned}
\end{equation}

\subsubsection{KK mode neutral NG boson $\boldsymbol\phi_y$ exchange}

\begin{equation}
\begin{aligned}
\begin{array}{c}
\includegraphics[scale=0.7]{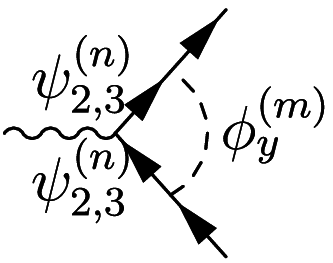}
%  \begin{picture}(0,0)(1,1)
%  \put(-42,12){$\psi_{2,3}^{(n)}$}
%  \put(-42,32){$\psi_{2,3}^{(n)}$}
%  \put(-5,22){$\phi_y^{(m)}$}
%  \end{picture}
\end{array}
=&
 \frac{e}{3} \dk4\dx 
 \frac{g_4^2Xm_n(-1)^{n+m}\gamma_5P_{\mu}}
 {[k^2-Xm_n^2-(1-X)M_m^2]^3}
 \\
&\times \frac{\pi RM}{\e^{2\pi RM}-1} \Bigg[
  \left(\frac{I_{4}}{\pi R}\right)^2
  \frac{m_n\tilde{m}_{nl}}{m_n^2-m_l^2}
  - \left(\frac{I_6}{\pi R}\right)^2
  \left(\frac{M_n}{m_n}\right)^2
  \frac{m_l\tilde{m}_{nl}}{m_n^2-m_l^2}
 \Bigg]. 
\end{aligned}
\end{equation}

\subsubsection{Zero mode neutral NG boson $\boldsymbol \phi_y$ exchange}

\begin{equation}
\begin{aligned}
\begin{array}{c}
\includegraphics[scale=0.7]{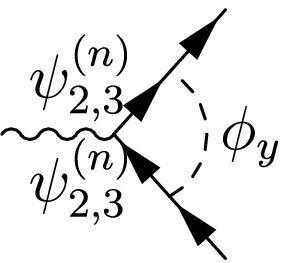}
%  \begin{picture}(0,0)(1,1)
%  \put(-42,12){$\psi_{2,3}^{(n)}$}
%  \put(-42,32){$\psi_{2,3}^{(n)}$}
%  \put(-5,22){$\phi_y$}
%  \end{picture}
\end{array}
=&
 -\frac{e}{3} \dk4\dx 
 \frac{g_4^2 I_5^2 \frac{m_l\tilde{m}_{nl}}{m_n^2-m_l^2}(-1)^nXm_n}
 {[k^2-Xm_n^2]^3}\gamma_5P_{\mu}. 
\end{aligned}
\end{equation}

\subsection{Charged current sector}

\subsubsection{KK mode W boson exchange}

%%%%%%%%%%%%%%%%%%%%%%%%%%%%%%%%%%%%%%%%%%%%%%%%%%%%%%%%
\def\da#1#2 {(#1)^{\dag}\times (#2)}
\def\epi{\frac{\pi RM}{\e^{2\pi RM}-1}}
\def\eip{\frac{\pi RM}{1-\e^{-2\pi RM}}}
%%%%%%%%%%%%%%%%%%%%%%%%%%%%%%%%%%%%%%%%%%%%%%%%%%%%%%%%

\begin{equation}
\begin{array}{c}
\includegraphics[scale=0.7]{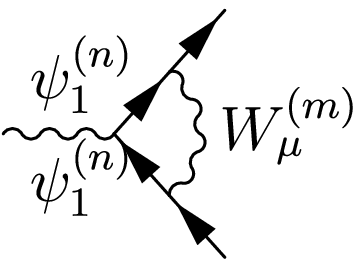}
%  \begin{picture}(0,0)(1,1)
%  \put(-42,12){$\psi_{1}^{(n)}$}
%  \put(-42,32){$\psi_{1}^{(n)}$}
%  \put(-5,22){$W_{\mu}^{(m)}$}
%  \end{picture}
\end{array}
=
 \frac{8}{3}e \dk4\dx
 \frac{g_4^2\frac{\hat{m}_l}{m_l}\sqrt{\epi}\frac{I_{4}}{\sqrt{\pi R}}
 \left(\frac{M_m}{m_n}I_1+I_{2}\right)
 (1-X)m_n}
 {[k^2-Xm_n^2-(1-X)M_m^2]^3}
 \gamma_5P_{\mu}. 
\end{equation}

\subsubsection{KK mode $X_{\mu}$ boson exchange}

\begin{equation}
\begin{array}{c}
\includegraphics[scale=0.7]{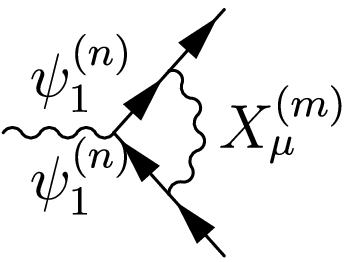}
%  \begin{picture}(0,0)(1,1)
%  \put(-42,12){$\psi_{1}^{(n)}$}
%  \put(-42,32){$\psi_{1}^{(n)}$}
%  \put(-5,22){$X_{\mu}^{(m)}$}
%  \end{picture}
\end{array}
=
 \frac{8}{3}e \dk4\dx
 \frac{g_4^2\frac{\hat{m}_l}{m_l}\sqrt{\epi}\frac{I_{4}}{\sqrt{\pi R}}
 \left(\frac{M_m}{m_n}I_1+I_{2}\right)
 (1-X)m_n}
 {[k^2-Xm_n^2-(1-X)M_m^2]^3}
 \gamma_5P_{\mu}. 
\end{equation}

\subsubsection{KK mode $X_y$ boson exchange}

\begin{equation}
\begin{array}{c}
\includegraphics[scale=0.7]{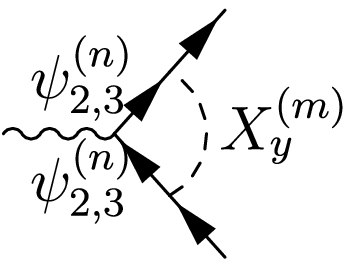}
%  \begin{picture}(0,0)(1,1)
%  \put(-42,12){$\psi_{2,3}^{(n)}$}
%  \put(-42,32){$\psi_{2,3}^{(n)}$}
%  \put(-5,22){$X_y^{(m)}$}
%  \end{picture}
\end{array}
=
 -\frac{2}{3}e\dk4\dx \frac{ -g_4^2\frac{\hat{m}_l}{m_l}\sqrt{\epi}
 \left((-1)^{l+m+n}\frac{I_{4}I_{2}}{\sqrt{\pi R}}+\frac{M_n}{m_n}\frac{I_1I_6}{\sqrt{\pi R}}\right)Xm_n}
 {[k^2-Xm_n^2-(1-X)M_m^2]^3}\gamma_5P_{\mu}. 
\end{equation}

\subsubsection{KK mode $W_y$ boson exchange}

\begin{equation}
\begin{array}{c}
\includegraphics[scale=0.7]{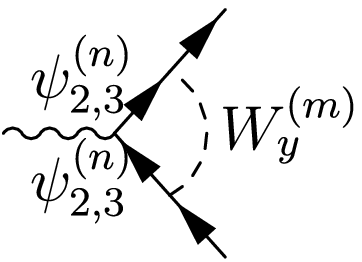}
%  \begin{picture}(0,0)(1,1)
%  \put(-42,12){$\psi_{2,3}^{(n)}$}
%  \put(-42,32){$\psi_{2,3}^{(n)}$}
%  \put(-5,22){$W_y^{(m)}$}
%  \end{picture}
\end{array}
=
 -\frac{2}{3}e \dk4\dx \frac{ -g_4^2\frac{\hat{m}_l}{m_l}\sqrt{\epi}
 \left((-1)^{l+m+n}\frac{I_{4}I_{2}}{\sqrt{\pi R}}+\frac{M_n}{m_n}\frac{I_1I_6}{\sqrt{\pi R}}\right)Xm_n}
 {[k^2-Xm_n^2-(1-X)M_m^2]^3}\gamma_5P_{\mu}. 
\end{equation}
%$\da{79}{79}$
%\begin{equation}
%N=0
%\end{equation}

\subsection{gauge boson self-energy}

\subsubsection{KK mode $W_{\mu}$ boson self energy diagram}

\begin{equation}
\begin{array}{c}
\includegraphics[scale=0.7]{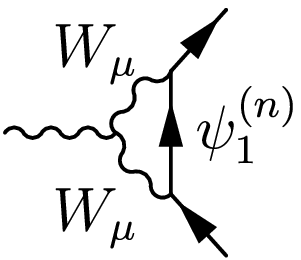}
%\begin{picture}(0,0)(1,1)
%\put(-10,22){$\psi_1^{(n)}$}
%\put(-38,38){$W_{\mu}$}
%\put(-38,6){$W_{\mu}$}
%\end{picture}
\end{array}
=
 e\dk4\dx
 \frac{g_4^2\frac{\hat{m}_l}{m_l}\sqrt{\epi}\frac{I_{4}}{\sqrt{\pi R}}
 \left(\frac{M_m}{m_n}I_1 +I_{2}\right)
 (6-3X)m_n}
 {[k^2-XM_m^2-(1-X)m_n^2]^3}\gamma_5P_{\mu}. 
\end{equation}

\subsubsection{KK mode $X_{\mu}$ boson self energy diagram}

\begin{equation}
\begin{array}{c}
\includegraphics[scale=0.7]{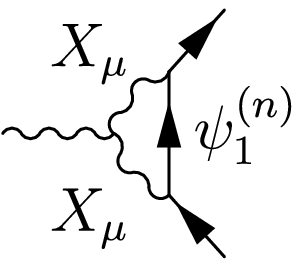}
%\begin{picture}(0,0)(1,1)
%\put(-10,22){$\psi_1^{(n)}$}
%\put(-38,38){$X_{\mu}$}
%\put(-38,6){$X_{\mu}$}
%\end{picture}
\end{array}
=
 e\dk4\dx
 \frac{g_4^2\frac{\hat{m}_l}{m_l}\sqrt{\epi}\frac{I_{4}}{\sqrt{\pi R}}
 \left(\frac{M_m}{m_n}I_1 +I_{2}\right)
 (6-3X)m_n}
 {[k^2-XM_m^2-(1-X)m_n^2]^3}\gamma_5P_{\mu}. 
\end{equation}

\subsubsection{KK mode $X_y$ boson self energy diagram} 

\begin{equation}
\begin{array}{c}
\includegraphics[scale=0.7]{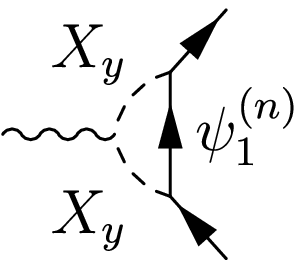}
%\begin{picture}(0,0)(1,1)
%\put(-10,22){$\psi_1^{(n)}$}
%\put(-38,38){$X_y$}
%\put(-38,6){$X_y$}
%\end{picture}
\end{array}
=
 -e\dk4\dx
 \frac{-g_4^2\frac{\hat{m}_l}{m_l}\sqrt{\epi}
 \left((-1)^{l+m+n} \frac{I_{4}I_{2}}{\sqrt{\pi R}}+\frac{M_n}{m_n}\frac{I_1I_6}{\sqrt{\pi R}}\right)
 (1-X)m_n
 }
 {[k^2-XM_m^2-(1-X)m_n^2]^3}\gamma_5P_{\mu}. 
\end{equation}

\subsubsection{KK mode $W_y$ boson self energy diagram}

\begin{equation}
\begin{array}{c}
\includegraphics[scale=0.7]{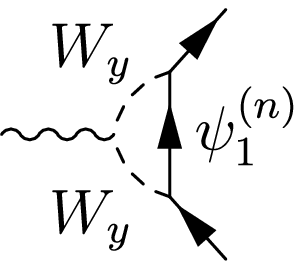}
%\begin{picture}(0,0)(1,1)
%\put(-10,22){$\psi_1^{(n)}$}
%\put(-38,38){$W_y$}
%\put(-38,6){$W_y$}
%\end{picture}
\end{array}
=
 e\dk4\dx
 \frac{g_4^2\frac{\hat{m}_l}{m_l}\sqrt{\epi}
 \left((-1)^{l+m+n}\frac{I_{4}I_{2}}{\sqrt{\pi R}}+\frac{M_n}{m_n}\frac{I_1I_6}{\sqrt{\pi R}}\right)
 (1-X)m_n}
 {[k^2-XM_m^2-(1-X)m_n^2]^3}\gamma_5P_{\mu}. 
\end{equation}
Various integrals appearing in the amplitudes are defined as follows. 
\begin{eqnarray}
I_1 &\equiv& \int_{-\pi R}^{\pi R} {\rm d}y \frac{1}{(\sqrt{\pi R})^3} S_l(y) C_m(y)S_n (y), \\
I_{2} &\equiv& \int_{-\pi R}^{\pi R} {\rm d}y \frac{1}{(\sqrt{\pi R})^3} \frac{M_n}{m_n} 
S_l(y) S_m(y) \left(C_n(y) - \frac{MR}{n} \varepsilon(y) S_n(y) \right), \\
I_{3} &\equiv& \int_{-\pi R}^{\pi R} {\rm d}y \e^{-M|y|} C_m(y) \left(C_n(y) - \frac{MR}{n} \varepsilon(y) S_n(y) \right), \\
I_{4} &\equiv& \int_{-\pi R}^{\pi R} {\rm d}y \e^{M|y|} S_m(y) S_n (y), \\
I_{5} &\equiv& \int_{-\pi R}^{\pi R} {\rm d}y \frac{M_n}{\sqrt{\pi R}m_n} \e^{-M|y|} \sqrt{\frac{M}{1-\e^{-2\pi MR}}} 
\left(C_n(y) + \frac{MR}{n} \varepsilon(y) S_n(y) \right), \\
I_{6} &\equiv& \int_{-\pi R}^{\pi R} {\rm d}y C_m(y) \left(C_n(y) - \frac{MR}{n} \varepsilon(y) S_n(y) \right) 
\end{eqnarray}
with $S_n(y) = \sin(\frac{n}{R}y)$ and $C_n(y) = \cos(\frac{n}{R}y)$. 

It is very interesting that the neutron EDM is generated already
at 1-loop although it is generated at 3-loop in the case of standard model.
We can see that the EDM contributions from neutral current sector are 
due to the mixing terms between different nonzero KK modes of down quark, which is proportional to $\tilde{m}_{nl}$. 
On the other hand, the EDM contributions from the charged current sector are 
due to the mixing terms between a zero mode and nonzero KK modes of down quark, 
which is proportional to $\hat{m}_n$. 
Since these two mass parameters $\tilde{m}_{nl}$ and $\hat{m}_n$ are proportional to 
both of the bulk mass $M$ and the W-boson mass $m_W$ (see, (\ref{mass})), 
the EDM vanishes if the bulk mass is zero while the Higgs VEV is nonzero,
and vice versa. 
This is consistent with the general discussion described in the introduction on 
how CP is violated.

\section{Numerical estimation of EDM from nonzero KK modes}

We move to numerical calculation of neutron electric dipole moment.
We expect that up quark electric dipole moment $d_u$ vanishes in this model, 
since there is no right-handed up quark and the operator describing up quark 
electric dipole moment $\langle H\rangle \bar{u}_{L(R)}\sigma_{\mu\nu}\gamma^5u_{R(L)}F^{\mu\nu}$ does not exist.
Thus, the neutron electric dipole moment $d_n$ in this model is written as follows:
\begin{equation}
d_n=\frac{4}{3}d_d -\frac{1}{3}d_u=\frac{4}{3}d_d.
\end{equation}

%\subsection{Input parameters}
To reproduce down quark Yukawa coupling, we must set a bulk mass parameter 
so as to satisfy the following relation:
\begin{equation}
\frac{m}{m_W}=\frac{2\pi RM}{\sqrt{(1-\e^{-2\pi RM})(\e^{2\pi RM}-1)}}
\sim \frac{4\sim 8{\rm MeV}}{80{\rm GeV}}. 
%=(0.5\sim 1)\times 10^{-4}.
\end{equation}
%However, the above condition can not be solved analytically, plotting function 
%$f(x)=\frac{x}{\sqrt{(1-\e^{-x})(\e^x -1)}}$
%and read off intersections of $f(x)$ and $x=5\times 10^{-5},1\times 10^{-4}$.
%The lines $1\times 10^{-4},5\times 10^{-5}$ in below graphs corresponds to 
% down quark mass 4MeV, 8MeV, respectively.
%\\
%\includegraphics[scale=0.35,angle=-90]{./graph/graph01.ps}
%\includegraphics[scale=0.35,angle=-90]{./graph/graph02.ps}\\
Thus, we set the bulk mass as $2\pi RM = 25.5$ ($m=6$ MeV is taken). 

Here, only the numerical results are shown. 
%and detailed calculation are summarized in Appendix. 
The contributions from the neutral current and the charged current 
processes to the EDM are denoted by $d(\rm N.C.)$ and $d(\rm C.C.)$ 
and are obtained as follows.
\begin{equation}
\begin{aligned}
 d({\rm N.C.})
%=&
% d(\gamma_{\mu})+d(\gamma_y)+d(h)+d(h_{\mu})
% +d(Z_{\mu})+ d(Z_y) + d(\phi_{\mu}) + d(\phi_y)
% \\
%=&
% \frac{4(MR)^4}{\pi^4}d_W
% \left[
% -\frac{4}{3}S_{\gamma_{\mu}}+S_{\gamma_y}-\frac{1}{3}S_h
%  -4(S_{h_{\mu}}^1+S_{h_{\mu}}^2)-4S_{Z_{\mu}}-S_{\phi_y}
% \right]
% \\
\sim&
 \frac{16}{9} e^3 \left(\frac{MR}{\pi}\right)^4 R^2 m_W(-8.3 \times 10^{-7}), \\
 d{(\rm C.C.)}
\sim&
 -\frac{2}{9}e^3 \frac{(MR)^3}{\pi^3}R^2 m_W (2.12\times 10^{-5})
 + \frac{16}{9}e^3 \frac{(MR)^4}{\pi^4} R^2m_W (8.0\times 10^{-7}). 
\end{aligned}
\end{equation}
Combining these results we obtain the final result on the contribution from nonzero KK mode
$d(\rm KK)$ as
\begin{equation}
\begin{aligned}
 d({\rm KK})
=&
 d({\rm N.C.})+d({\rm C.C.})
% \nt\\
%=&
%  -\frac{(MR)^3}{2\pi^3}d_W\cdot 2.12\times 10^{-5}
%  +\frac{4(MR)^4}{\pi^4}d_W(\cdot 8.0\times 10^{-7}-8.3\times 10^{-7})
%  \nt\\
%=&
% -2.3\times 10^{-5}\cdot \underbrace{d_W}_{7.073\times 10^{-21}(Rm_W)^2}
% [e_l\cdot{\rm cm}]
% \nt\\
\sim
 -2.3 \times 10^{-23}(Rm_W)^2[e \cdot{\rm cm}].
\end{aligned}
\end{equation}
We require that the contribution of KK mode $\frac{4}{3}d(\rm KK)$ 
is less than the experimental upper bound \cite{PDG}, 
\begin{equation}
\frac{4}{3}\cdot 2.3 \times 10^{-23}(Rm_W)^2 [e \cdot{\rm cm}]< 2.9\times 10^{-26}[e \cdot{\rm cm}]
\end{equation}
which gives a lower bound for the compactification scale
\begin{equation}
\frac{1}{R}>
%\sqrt{\frac{64}{8.7}}m_W\sim 
33m_W \simeq 2.6~{\rm TeV}. 
\end{equation}

\section{Summary}

In this paper we studied the neutron electric dipole moment (EDM) in a five dimensional 
$SU(3)$ gauge-Higgs unification compactified on $M^4\times S^1/Z_2$ space-time
including massive fermions belonging to the triplet of $SU(3)$. 
The smallness of the quark Yukawa coupling is realized 
by introducing a $Z_{2}$-odd bulk mass $M\epsilon(y)$.
We pointed out that to realize the CP violation 
is a non-trivial task in the gauge-Higgs unification scenario 
where the Yukawa coupling is originally gauge coupling, which is of course real. 
We identified the transformation properties of each field 
under P and CP transformations, 
since to get non-vanishing EDM both P and CP symmetries have to be broken,
though P is broken anyway by the orbifolding. 
We have found that since the theory itself is CP symmetric,
the unique source of the CP violation in our model is the VEV of the Higgs,
the extra space component of the gauge field $A_{y}$.
In such a sense, CP is broken spontaneously through the Hosotani mechanism\cite{Hosotani}.
We emphasized that actually to get physically the CP violating effect the
interplay between the VEV of $A_y$ and the bulk mass $M$ is crucial.
In fact in the hypothetical limit of $M \to 0$,
it turned out that by suitable chiral transformation $A_y$ becomes
a field with even CP eigenvalue, whose VEV therefore does not break CP symmetry. 
From such a point of view both of the VEV and the bulk mass 
are the cause of the CP violation on an equal footing. 

We then calculated the one-loop contributions to 
the neutron electric dipole moment as the typical example of the
CP violating observable and found that the EDM appears already at the one-loop level, 
without invoking to the three generation scheme, 
in clear contrast to the case of the Standard Model where EDM appears
only at the three-loop level.
The explicit calculation has shown that one-loop contributions from nonzero
KK modes to the neutron EDM are generated due to the mixing effects 
between different nonzero KK modes, and between a zero mode and nonzero KK modes. 
Also, the obtained EDM was proportional to the Higgs VEV and the bulk mass,
which was consistent with what we discussed concerning
the importance of their interplay to get CP violation. 

Furthermore, we could confirm that the standard model contribution
to the neutron EDM due to the Kaluza-Klein zero modes vanishes
at the one-loop level, as we expect. 

The fact that EDM appears already at the one-loop level suggests that
we may be able to get a rather stringent lower bound on the compactification
scale by the comparison with the data.
This turns out to be the case. We could derive a rather stringent lower bound
for the compactification scale, 
which is around 2.6 TeV, by comparing the contribution due to the nonzero Kaluza-Klein modes with the experimental data.  

\subsection*{Acknowledgments}

The work of the authors was supported 
in part by the Grant-in-Aid for Scientific Research 
of the Ministry of Education, Science and Culture, No.18204024 and No. 20025005.  

%\appendix


\begin{thebibliography}{n}


%%%%
\bibitem{Manton} 
  N.~S.~Manton,
  %``A New Six-Dimensional Approach To The Weinberg-Salam Model,''
  Nucl.\ Phys.\ B {\bf 158}, 141 (1979).

\bibitem{Fairlie}
  D.~B.~Fairlie,
  %``Higgs' Fields And The Determination Of The Weinberg Angle,''
  Phys.\ Lett.\ B {\bf 82}, 97 (1979); 
  %``Two Consistent Calculations Of The Weinberg Angle,''
  J.\ Phys.\ G {\bf 5}, L55 (1979).


%%%%%%%%%%%%%%%%%%%%%%%%%%%%%%%
\bibitem{Hosotani}  
  Y.~Hosotani,
    %``Dynamical Mass Generation By Compact Extra Dimensions,''
  Phys.\ Lett.\ B {\bf 126}, 309 (1983); 
  %``Dynamical Gauge Symmetry Breaking As The Casimir Effect,''
  Phys.\ Lett.\ B {\bf 129}, 193 (1983); 
  %``DYNAMICS OF NONINTEGRABLE PHASES AND GAUGE SYMMETRY BREAKING,''
  Annals Phys.\  {\bf 190}, 233 (1989).

%%%%%%%%%%%%%%% 
\bibitem{Antoniadis}
 % A Possible new dimension at a few TeV.
 I. Antoniadis, Phys. Lett. B {\bf 246}, 377-384 (1990). 



%%%%%%%%%%%%%%%%%%%%%%%%%%%%%%%%%%%%%%%%%%%%%%%%%%%%%%%%%%%%%%% 
\bibitem{HIL}
  H.~Hatanaka, T.~Inami and C.~S.~Lim,
  %``The gauge hierarchy problem and higher dimensional gauge theories,''
  Mod.\ Phys.\ Lett.\ A {\bf 13}, 2601 (1998). 
%  [arXiv:hep-th/9805067].
%%%%%%%%%%%%%%%%%%%%%%%%%%%%%%%%%%%%%%%%%%%%%%%%%%%%%%%%%%%%%%



%%%%% 
\bibitem{KLY}  
  M.~Kubo, C.~S.~Lim and H.~Yamashita,
  %``The Hosotani mechanism in bulk gauge theories with an orbifold 
  %extra  space S(1)/Z(2),''
  Mod.\ Phys.\ Lett.\ A {\bf 17}, 2249 (2002). 
%  [arXiv:hep-ph/0111327].


%%%%%%%%%%%%%%% 
%%%%% 
\bibitem{BN}
  G.~Burdman and Y.~Nomura,
  %``Unification of Higgs and gauge fields in five dimensions,''
  Nucl.\ Phys.\ B {\bf 656}, 3 (2003). 
%  [arXiv:hep-ph/0210257].

\bibitem{CGM}
  C.~Cs\'aki, C.~Grojean and H.~Murayama,
  %``Standard model Higgs from higher dimensional gauge fields,''
  Phys.\ Rev.\ D {\bf 67}, 085012 (2003). 
%  [arXiv:hep-ph/0210133].

\bibitem{GMN}
  I.~Gogoladze, Y.~Mimura and S.~Nandi,
  %``Gauge Higgs unification on the left-right model,''
  Phys.\ Lett.\ B {\bf 560}, 204 (2003); 
%  [arXiv:hep-ph/0301014]; 
  %``Coupling unifications in gauge-Higgs unified orbifold models,''
  Phys.\ Rev.\ D {\bf 72}, 055006 (2005).  
  %[arXiv:hep-ph/0504082], 

\bibitem{SSS}
C.~A.~Scrucca, M.~Serone and L.~Silvestrini,
  %``Electroweak symmetry breaking and fermion masses from extra dimensions,''
  Nucl.\ Phys.\ B {\bf 669}, 128 (2003).  %[arXiv:hep-ph/0304220]; 

\bibitem{HHKY}
 N.~Haba, Y.~Hosotani, Y.~Kawamura and T.~Yamashita,
  %``Dynamical symmetry breaking in gauge-Higgs unification on orbifold,''
  Phys.\ Rev.\ D {\bf 70}, 015010 (2004).  
  %[arXiv:hep-ph/0401183]

\bibitem{HNT}
Y.~Hosotani, S.~Noda and K.~Takenaga,
  %``Dynamical gauge symmetry breaking and mass generation on the orbifold
  %T**2/Z(2),''
  Phys.\ Rev.\ D {\bf 69}, 125014 (2004); 
%  [arXiv:hep-ph/0403106]
  %``Dynamical gauge-Higgs unification in the electroweak theory,''
  Phys.\ Lett.\ B {\bf 607}, 276 (2005).  
%  [arXiv:hep-ph/0410193].
  

\bibitem{MSSS}
G.~Martinelli, M.~Salvatori, C.~A.~Scrucca and L.~Silvestrini,
  %``Minimal gauge-Higgs unification with a flavour symmetry,''
  JHEP {\bf 0510}, 037 (2005). % [arXiv:hep-ph/0503179]. 

%\bibitem{HMOY}
% N.~Haba, S.~Matsumoto, N.~Okada and T.~Yamashita,
  %``Effective theoretical approach of gauge-Higgs unification model and its
  %phenomenological applications,''
%  JHEP {\bf 0602}, 073 (2006);
  %[hep-ph/0511046]
%    I.~Gogoladze, N.~Okada and Q.~Shafi,
  %``Higgs Boson Mass From Gauge-Higgs Unification,''
%  Phys.\ Lett.\  B {\bf 655}, 257 (2007); 
% [arXiv:0705.3035 [hep-ph]].
%  I.~Gogoladze, N.~Okada and Q.~Shafi,
  %``Window For Higgs Boson Mass From Gauge-Higgs Unification,''
%  Phys.\ Lett.\  B {\bf 659}, 316 (2008); 
%  [arXiv:0708.2503 [hep-ph]].
%    N.~Haba, S.~Matsumoto, N.~Okada and T.~Yamashita,
  %``Effective Potential of Higgs Field in Warped Gauge-Higgs Unification,''
%  Prog.\ Theor.\ Phys.\  {\bf 120}, 77 (2008). 
%  [arXiv:0802.3431 [hep-ph]].


\bibitem{BQ}
  C.~Biggio and M.~Quiros,
  %``Higgs-gauge unification without tadpoles,''
  Nucl.\ Phys.\ B {\bf 703}, 199 (2004). 
%  [arXiv:hep-ph/0407348].

\bibitem{PS}
G.~Panico and M.~Serone,
  %``The electroweak phase transition on orbifolds with gauge-Higgs
  %unification,''
  JHEP {\bf 0505}, 024 (2005). %[arXiv:hep-ph/0502255]. 


\bibitem{CCP}
G.~Cacciapaglia, C.~Csaki and S.~C.~Park,
  %``Fully radiative electroweak symmetry breaking,''
  JHEP {\bf 0603}, 099 (2006). %arXiv:hep-ph/0510366. 

\bibitem{PSW}
G.~Panico, M.~Serone and A.~Wulzer,
  %``A model of electroweak symmetry breaking from a fifth dimension,''
  Nucl.\ Phys.\ B {\bf 739}, 186 (2006); %arXiv:hep-ph/0510373; 
  %``Electroweak symmetry breaking and precision tests with a fifth
  %dimension,''
  Nucl.\ Phys.\ B {\bf 762}, 189 (2007). 
%  [arXiv:hep-ph/0605292].

%%%%%%%%%%%%%%% 

\bibitem{MT}
  N.~Maru and K.~Takenaga,
  %``Aspects of phase transition in gauge-Higgs unification at finite
  %temperature,''
  Phys.\ Rev.\ D {\bf 72}, 046003 (2005);   
%  [arXiv:hep-th/0505066]; 
  %``Effects of bulk mass in gauge-Higgs unification,''
  Phys.\ Lett.\ B {\bf 637}, 287 (2006); 
%  [arXiv:hep-ph/0602149].   
  %``Bulk mass effects in gauge-Higgs unification at finite temperature,''
  Phys.\ Rev.\ D {\bf 74}, 015017 (2006). 
%  [arXiv:hep-ph/0606139]. 


\bibitem{LM}
  C.~S.~Lim and N.~Maru,
  %``Calculable One-Loop Contributions to S and T Parameters in the
  %Gauge-Higgs Unification,''
  Phys.\ Rev.\  D {\bf 75}, 115011 (2007). 
%  [arXiv:hep-ph/0703017].



\bibitem{ACP}
K.~Agashe, R.~Contino and A.~Pomarol,
  %``The minimal composite Higgs model,''
  Nucl.\ Phys.\ B {\bf 719}, 165 (2005);  
%  [arXiv:hep-ph/0412089]; 
  K.~Agashe and R.~Contino,
  %``The minimal composite Higgs model and electroweak precision tests,''
  Nucl.\ Phys.\ B {\bf 742}, 59 (2006); 
%  [arXiv:hep-ph/0510164].
  K.~Agashe, R.~Contino, L.~Da Rold and A.~Pomarol,
  %``A custodial symmetry for Z b anti-b,''
  Phys.\ Lett.\ B {\bf 641}, 62 (2006); 
%  [arXiv:hep-ph/0605341].
  R.~Contino, L.~Da Rold and A.~Pomarol,
  %``Light custodians in natural composite Higgs models,''
    Phys.\ Rev.\  D {\bf 75}, 055014 (2007). 
%  arXiv:hep-ph/0612048.



\bibitem{OW}
  K.~y.~Oda and A.~Weiler,
  %``Wilson lines in warped space: Dynamical symmetry breaking and
  %restoration,''
  Phys.\ Lett.\ B {\bf 606}, 408 (2005).  
  %[arXiv:hep-ph/0410061]. 

\bibitem{HM}
 Y.~Hosotani and M.~Mabe,
  %``Higgs boson mass and electroweak-gravity hierarchy from dynamical
  %gauge-Higgs unification in the warped spacetime,''
  Phys.\ Lett.\ B {\bf 615}, 257 (2005); 
  %[arXiv:hep-ph/0503020]. 
Y.~Hosotani, S.~Noda, Y.~Sakamura and S.~Shimasaki,
  %``Gauge-Higgs unification and quark-lepton phenomenology in the warped
  %spacetime,'' 
  Phys.\ Rev.\ D {\bf 73}, 096006 (2006); 
  %[arXiv:hep-ph/0601241]. 
  Y.~Sakamura and Y.~Hosotani,
  %``W W Z, W W H, and Z Z H couplings in the dynamical gauge-Higgs  
  %unification in the warped spacetime,''
  Prog.\ Theor.\ Phys.\  {\bf 118}, 935 (2007); 
  %[arXiv:hep-ph/0703212]. 
    Y.~Hosotani, K.~Oda, T.~Ohnuma and Y.~Sakamura,
  %``Dynamical Electroweak Symmetry Breaking in SO(5)xU(1) Gauge-Higgs
  %Unification with Top and Bottom Quarks,''
  Phys.\ Rev.\  D {\bf 78}, 096002 (2008). 
%  [arXiv:0806.0480 [hep-ph]].

\bibitem{CPSW}
  M.~Carena, E.~Ponton, J.~Santiago and C.~E.~M.~Wagner,
  %``Light Kaluza-Klein states in Randall-Sundrum models with custodial
  %SU(2),''
  Nucl.\ Phys.\ B {\bf 759}, 202 (2006); 
%  [arXiv:hep-ph/0607106]; 
  %``Electroweak constraints on warped models with custodial symmetry,''
    Phys.\ Rev.\  D {\bf 76}, 035006 (2007); 
%  [arXiv:hep-ph/0701055]
    A.~D.~Medina, N.~R.~Shah and C.~E.~M.~Wagner,
  %``Gauge-Higgs Unification and Radiative Electroweak Symmetry Breaking in
  %Warped Extra Dimensions,''
  Phys.\ Rev.\  D {\bf 76}, 095010 (2007); 
%  [arXiv:0706.1281 [hep-ph]].
  M.~Carena, A.~D.~Medina, B.~Panes, N.~R.~Shah and C.~E.~M.~Wagner,
  %``Collider Phenomenology of Gauge-Higgs Unification Scenarios in Warped Extra
  %Dimensions,''
  Phys.\ Rev.\  D {\bf 77}, 076003 (2008);  
%  [arXiv:0712.0095 [hep-ph]].
  M.~Carena, A.~D.~Medina, N.~R.~Shah and C.~E.~M.~Wagner,
  %``Gauge-Higgs Unification, Neutrino Masses and Dark Matter in Warped Extra
  %Dimensions,''
  arXiv:0901.0609 [hep-ph].

\bibitem{Hatanaka}
  H.~Hatanaka,
  %``Radiatively Induced Spontaneous Symmetry Breaking by Wilson Line in a
  %Warped Extra Dimension,''
  arXiv:0712.1334 [hep-th].


\bibitem{GUT}
  C.~S.~Lim and N.~Maru,
  %``Towards A Realistic Grand Gauge-Higgs Unification,''
  Phys.\ Lett.\  B {\bf 653}, 320 (2007). 
%  [arXiv:0706.1397 [hep-ph]].


\bibitem{FLHC}
  A.~Falkowski,
  %``Pseudo-Goldstone Higgs Production via Gluon Fusion,''
  Phys.\ Rev.\  D {\bf 77}, 055018 (2008). 
%  [arXiv:0711.0828 [hep-ph]].

\bibitem{LHC}
  N.~Maru and N.~Okada,
  %``Gauge-Higgs Unification at LHC,''
  Phys.\ Rev.\  D {\bf 77}, 055010 (2008); 
%  [arXiv:0711.2589 [hep-ph]].
  N.~Maru,
  %``Finite Gluon Fusion Amplitude in the Gauge-Higgs Unification,''
  Mod.\ Phys.\ Lett.\  A {\bf 23}, 2737 (2008).
%  [arXiv:0803.0380 [hep-ph]].


\bibitem{LMgy}
  C.~S.~Lim and N.~Maru,
  %``Calculable Violation of Gauge-Yukawa Universality and Top Quark Mass in the
  %Gauge-Higgs Unification,''
  arXiv:0904.0304 [hep-ph].
%%%%%%%%%%%%%%%%%%%%%%%%%%%%%%%%%%%%%%%%

\bibitem{HSY}
  N.~Haba, Y.~Sakamura and T.~Yamashita,
  %``Weak boson scattering in Gauge-Higgs Unification,''
  arXiv:0904.3177 [hep-ph].

\bibitem{ABQ}
  I.~Antoniadis, K.~Benakli and M.~Quiros,
  %``Finite Higgs mass without supersymmetry,''
  New J.\ Phys.\  {\bf 3}, 20 (2001). 
%  [arXiv:hep-th/0108005].


\bibitem{GIQ}
  G.~von Gersdorff, N.~Irges and M.~Quiros,
  %``Bulk and brane radiative effects in gauge theories on orbifolds,''
  Nucl.\ Phys.\ B {\bf 635}, 127 (2002). 
%  [arXiv:hep-th/0204223].

\bibitem{CNP}
 R.~Contino, Y.~Nomura and A.~Pomarol,
  %``Higgs as a holographic pseudo-Goldstone boson,''
  Nucl.\ Phys.\ B {\bf 671}, 148 (2003). 
%  [arXiv:hep-ph/0306259].
 
 
\bibitem{LMH}
  C.~S.~Lim, N.~Maru and K.~Hasegawa,
  %``Six dimensional gauge-Higgs unification with an extra space S**2 and the
  %hierarchy problem,''
    J.\ Phys.\ Soc.\ Jap.\  {\bf 77}, 074101 (2008). 
%  arXiv:hep-th/0605180.

\bibitem{HLM}
  K.~Hasegawa, C.~S.~Lim and N.~Maru,
  %``An attempt to solve the hierarchy problem based on gravity gauge Higgs
  %unification scenario,''
  Phys.\ Lett.\ B {\bf 604}, 133 (2004). 
%  [arXiv:hep-ph/0408028].

\bibitem{MY}
  N.~Maru and T.~Yamashita,
  %``Two-loop calculation of Higgs mass in gauge-Higgs unification: 
  %5D  massless QED compactified on S**1,''
  Nucl.\ Phys.\ B {\bf 754}, 127 (2006); 
%  [arXiv:hep-ph/0603237].
%\bibitem{HMTY}
  Y.~Hosotani, N.~Maru, K.~Takenaga and T.~Yamashita,
  %``Two loop finiteness of Higgs mass and potential in the gauge-Higgs
  %unification,''
  Prog.\ Theor.\ Phys.\  {\bf 118}, 1053 (2007). 
%  [arXiv:0709.2844 [hep-ph]].

%\bibitem{PT}
% M.~E.~Peskin and T.~Takeuchi,
  %``A New constraint on a strongly interacting Higgs sector,''
%  Phys.\ Rev.\ Lett.\  {\bf 65}, 964 (1990); 
  %``Estimation of oblique electroweak corrections,''
%  Phys.\ Rev.\ D {\bf 46}, 381 (1992).


\bibitem{ALM1}
  Y.~Adachi, C.~S.~Lim and N.~Maru,
  %``Finite Anomalous Magnetic Moment in the Gauge-Higgs Unification,''
  Phys.\ Rev.\  D {\bf 76}, 075009 (2007). 
%  [arXiv:0707.1735 [hep-ph]].

\bibitem{g-2RS}
  H.~Davoudiasl, J.~L.~Hewett and T.~G.~Rizzo,
  %``The (g-2) of the muon in localized gravity models,''
  Phys.\ Lett.\  B {\bf 493}, 135 (2000). 
%  [arXiv:hep-ph/0006097].


\bibitem{g-2UED}
  K.~Agashe, N.~G.~Deshpande and G.~H.~Wu,
  %``Can extra dimensions accessible to the SM explain the recent  measurement
  %of anomalous magnetic moment of the muon?,''
  Phys.\ Lett.\  B {\bf 511}, 85 (2001); 
%  [arXiv:hep-ph/0103235].
  T.~Appelquist and B.~A.~Dobrescu,
  %``Universal extra dimensions and the muon magnetic moment,''
  Phys.\ Lett.\  B {\bf 516}, 85 (2001). 
%  [arXiv:hep-ph/0106140].

\bibitem{ALM2}
  Y.~Adachi, C.~S.~Lim and N.~Maru,
  %``More on the Finiteness of Anomalous Magnetic Moment in the Gauge-Higgs
  %Unification,''
  Phys.\ Rev.\  D {\bf 79}, 075018 (2009).

\bibitem{ALM3}
  Y.~Adachi, C.~S.~Lim and N.~Maru,
  %``Lower Bound for Compactification Scale from Muon $g - 2$ in the Gauge-Higgs
  %Unification,''
  arXiv:0904.1695 [hep-ph]. 

\bibitem{Lim} 
  C.~S.~Lim , 
  Phys. Lett.\ B {\bf 256}, 233 (1991); 
  A.~Strominger and E.~Witten, 
  Commun. Math. Phys. {\bf 101}, 341 (1985). 

\bibitem{Poland} 
  B.~Grzadkowski and J.~Wudka, 
  Phys.Rev.Lett. {\bf 93}, 211603 (2004). 

%\bibitem{Schwinger}
%  J.~S.~Schwinger, Phys.\ Rev.\ {\bf 73}, 416 (1948). 
\bibitem{Shabalin}
  E.~P.~Shabalin,
  %``Electric Dipole Moment Of Quark In A Gauge Theory With Left-Handed
  %Currents,''
  Sov.\ J.\ Nucl.\ Phys.\  {\bf 28}, 75 (1978)
  [Yad.\ Fiz.\  {\bf 28}, 151 (1978)].

\bibitem{PDG}
  C.~Amsler et al. (Particle Data Group), Phys.\ Lett.\  B {\bf 667}, 1 (2008). 

\bibitem{Wudka}
B. Grzadkowski, J. Wudka, Phys.Rev.Lett. {\bf 97}, 211602 (2006). 
\end{thebibliography}
\end{document}